\documentclass{aastex62}

\graphicspath{{./}{figures/}}

\received{}
\revised{May 10, 2020}
\accepted{May 30, 2020}
\submitjournal{ApJ}

\shorttitle{Revisiting the Formation of HeH$^+$}
\shortauthors{Forrey et al.}

\begin{document}

\title{Revisiting the Formation of HeH$^+$ in the Planetary Nebula NGC 7027}

\correspondingauthor{P. C. Stancil}
\email{pstancil@uga.edu}

\author[0000-0002-0786-7307]{R. C. Forrey}
\affiliation{Department of Physics, Penn State University, Berks Campus, Reading, PA 19610}

\author{J. F. Babb}
\affiliation{Institute for Theoretical Atomic, Molecular, and Optical Physics, Harvard-Smithsonian Center for Astrophysics, Cambridge, MA 02138}

\author{E. D. S. Courtney}
\author{R. McArdle}
\author{P. C. Stancil}
\affiliation{Department of Physics and Astronomy and the Center for Simulational Physics, University of Georgia, Athens, GA 30602}



\begin{abstract}

From four independent calculations using three different theoretical approaches, rate coefficients for the formation of HeH$^+$ via the radiative association of He$^+$ and H were computed. Good agreement is found between our new calculations and prior results obtained two decades ago for kinetic temperatures between $\sim$800 and 20,000 K. This finding is inconsistent with a recent claim in the literature of a wide variation in published values and establishes the robustness of our knowledge of this process for the formation of HeH$^+$. The implications of the current results to the first detection of HeH$^+$ and its modeled abundance in the planetary nebula NGC 7027
are discussed.
\end{abstract}



\section{Introduction}

Bound molecular complexes involving helium are extremely rare,  but early laboratory studies of electron impact ionization of a H$_2$/He gas mixture suggested the presence of the helium-hydride cation HeH$^+$ \citep{hog25}. Immediately after the spectroscopic identification of HeH$^+$ in the laboratory \citep{dab78}, astronomers postulated its presence in the interstellar medium and developed chemical schemes for its formation \citep{bla78,flo79,rob82}. A prime environment for the presence of HeH$^+$ was predicted to be near the Str\"omgren radius of a planetary nebula (PN) with its formation dominated by the radiative association process,
\begin{equation}
    {\rm He}^+ + {\rm H} \rightarrow {\rm HeH}^+ + {\rm h}\nu.
    \label{he+hra}
\end{equation}
At about the same time, HeH$^+$ was proposed to play a role in the formation of H$_2^+$ and H$_2$ during the epoch of recombination in the early Universe through the sequence of reactions \citep{bla78,lep84},
\begin{equation}
    {\rm H}^+ + {\rm He} \rightarrow {\rm HeH}^+ + {\rm h}\nu,
    \label{heh+ra}
\end{equation}
\begin{equation}
    {\rm HeH}^+ + {\rm H} \rightarrow {\rm H}^+_2 + {\rm He},
    \label{heh+rea}
\end{equation}
and
\begin{equation}
    {\rm H}^+_2 + {\rm H} \rightarrow {\rm H}_2 + {\rm H}^+
    \label{h2form}
\end{equation}
\citep[for detailed models see][]{gal98,sta98}. Over the next four decades, astronomers diligently searched for astrophysical signatures of HeH$^+$ in PNe \citep{moo88,liu97,din01}, supernova ejecta \citep{mil92}, and high-redshift
quasars \citep{zin11}. While hints of its presence were tantalizing, only upper limits to its abundance could be obtained, leaving the existence of HeH$^+$ in astrophysical sources in question. 
Finally, \citet{gus19} made an unambiguous detection of the $J = 1 \rightarrow 0$ rotational transition line of HeH$^+$ in the PN NGC 7027. This was enabled by the development of a high spectral-resolution detector, termed the German Receiver for Astronomy at Terahertz Frequencies (GREAT), and its subsequent deployment on the Stratospheric Observatory for Infrared Astronomy (SOFIA). In particular, this detection was made possible by the ability of GREAT to separate the HeH$^+$ emission line from the hyperfine components of a nearby blended CH feature, which had plagued past efforts. This detection was a important advance with a number of significant implications for astrochemistry. However, modeling the intensity of the HeH$^+$ feature proved to be problematic. 
A comprehensive non-equilibrium chemical and level population model predicted the HeH$^+$ line intensity to be a factor of four smaller than the observation. \citet{gus19} argued that this discrepancy could be remedied by  scaling the literature values of the rate coefficient for process~(\ref{he+hra}) by a constant value.

More recently, \citet{neu20} observed the 1-0 P(1) and P(2) rovibrational lines of HeH$^+$ in NGC 7027. These emission features appear in the near infrared and were obtained with the iSHELL
spectrograph on NASA's Infrared Telescope Facility confirming the discovery of \citet{gus19}.

In this work, we demonstrate that the prior rate coefficients computed by other authors are robust and that an arbitrary scaling is unjustified. Further, using more modern theoretical approaches and molecular data, we provide updated rate coefficients for the radiative association process (\ref{he+hra}) with an estimated
uncertainty. We conclude by discussing possible resolutions to the observed and predicted HeH$^+$ line intensity, though a detailed model of NGC 7027 is beyond the scope of this work.

\section{Molecular Data}

Following the work on HeH$^+$ photodissociation of
\citet{miy12}, the ab initio potential energies for the ground X~$^1\Sigma^+$ and excited A~$^1\Sigma^+$ electronic states are taken from
\citet{bis79} and \citet{kra95}, respectively. Extensions of the ab initio potentials to both short- and long-range were performed as described in \citet{miy12}. The A~$^1\Sigma^+ \leftarrow$ X~$^1\Sigma^+$ transition dipole moment (TDM) function was adopted from \citet{kra95}
with short- and long-range extrapolations as given in
\citet{miy12}. The accuracies of the potential energies are discussed in the respective manuscripts, while the photodissociation cross section computed by \citet{miy12} was found to be consistent with the only available experimental value, suggesting that the TDM is reliable. For the ground electronic state, we computed 162 bound rovibrational states consistent with the calculations of \citet{zyg98}.

\section{Radiative Association Theory}

In this work, we utilized two different theoretical approaches to
compute the cross sections and rate coefficients for the radiative
association process (\ref{he+hra}). In the first method, we used the
standard two-state quantum-mechanical Fermi Golden Rule (FGR) approach which should be valid in the zero-density limit (ZDL), henceforth referred to as FGR-ZDL. The FGR-ZDL cross section is given by
\begin{equation}
\sigma(E) = \sum_{J^\prime}\sum_{v^{\prime\prime}} \sigma_{J^\prime}(v^{\prime\prime},E),   
\label{sigsum}
\end{equation}
where the partial cross section is given by
\begin{equation}
    \sigma_{J^\prime}(v^{\prime\prime},E) =\frac{64}{3}\frac{\pi^5}{c^3}\frac{\nu^3}{k^2}p \bigl [J^\prime M^2_{v^{\prime\prime},J^\prime-1;k,J^\prime} +
    (J^\prime+1)M^2_{v^{\prime\prime},J^\prime+1;k,J^\prime} \bigr ],
\end{equation}
and $\nu$ is the photon frequency, $k$ the initial wave vector, and $p$ the approach probability factor in the initial electronic state, which is 1/4 in this case\footnote{We note that the factor of 1/4 was neglected in earlier calculations, but all comparisons are corrected for this factor here.}. $J^\prime$ is the initial rotational quantum number, while $v^{\prime\prime}, J^{\prime\prime}$ is the vibrational and rotational quantum numbers for the final product HeH$^+$, and $M$ is the electric TDM element between the initial
and final states.

In the second approach,
a kinetic model was adopted which accounts for both direct and indirect radiative association processes \citep{for13,for15}. This method is applied to
two limiting cases in which the gas is taken to be in local thermodynamic equilibrium (LTE) and in non-LTE (NLTE) in the ZDL; these cases are expected to be valid at high and low
densities, respectively, with typical astrophysical conditions lying between the two
limits. Radiative association results in the two limits will be referred to as LTE and NLTE-ZDL, respectively. 
It has been shown in our earlier work that LTE and NLTE-ZDL rate coefficients merge at relatively high temperatures, while the NLTE-ZDL and FGR-ZDL results are nearly identical for all temperatures 
\citep[see for example][]{cai17}.

\section{Results}

\begin{figure}
\plotone{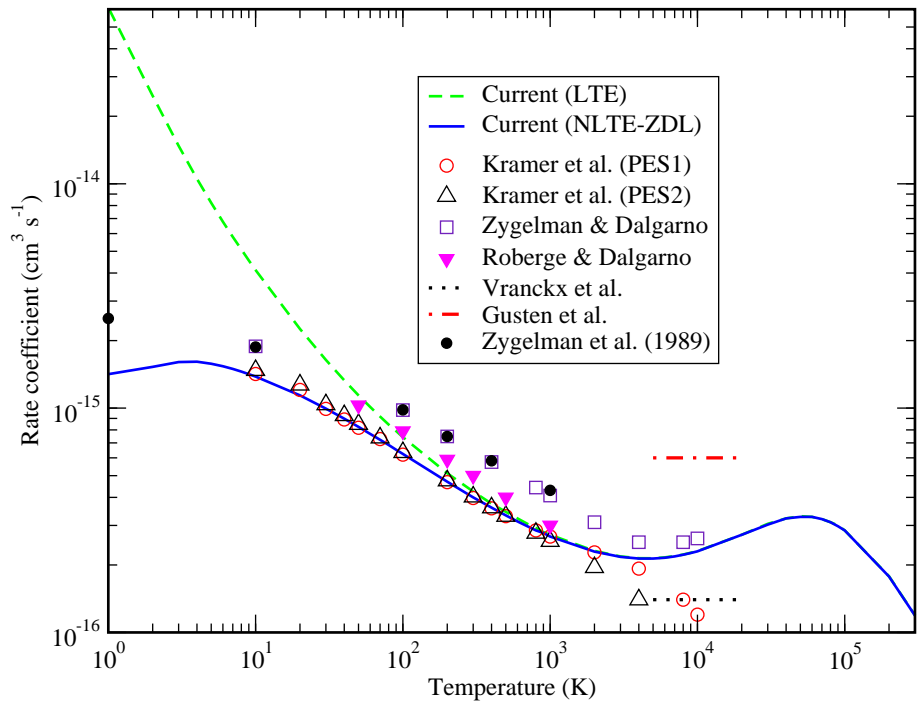}
\caption{Comparison of the He$^+$ + H radiative association rate coefficients. Sources as indicated in the legend. The values indicated as Vranckx et al. were deduced by \citet{gus19} from the cross sections of \citet{vra13}.}
\label{fig1}
\end{figure}

Using the LTE, NLTE-ZDL, and FGR-ZDL methods, we computed total
cross sections and rate coefficients for process~(\ref{he+hra}).
Figure~\ref{fig1} displays the resulting rate coefficients as a
function of kinetic temperature.\footnote{Total and final rovibrational-state-resolved cross sections and rate coefficients, stimulated radiative
association results, and further details are given in \citet{cou20}.} The current LTE and NLTE-ZDL results are seen to
converge for temperatures above $\sim$300~K. Our FGR-ZDL results are not shown as
they are indistinguishable from the NLTE-ZDL rate coefficients on the
scale of the figure. It should also be noted that two independent
FGR-ZDL codes were utilized in the calculations and gave consistent
results. 

The current rate coefficients are compared to prior calculations
using the FGR-ZDL method \citep{zyg90,kra95}, a quantum decay
method \citep{zyg89}, and approximate results
\citep{san71} as reported in \citet{rob82}. The 
results of \citet{kra95}, which adopted two different sets
of ab initio potentials, are in excellent agreement with the current
rate coefficients for $T<2000$~K. The results of \citet{rob82},
\citet{zyg89}, and \citet{zyg90} are slightly larger than the
current FGR-ZDL and NLTE-ZDL calculations, but follow the
same temperature dependence. This difference is likely related to the adoption of different TDM functions and potential
energies, while the difference at high temperatures with the
\citet{kra95} results may be due to convergence errors in
the latter (e.g., the maximum kinetic energy in their cross sections occur near 10,000 cm$^{-1}$, which would result in an underestimation of the rate coefficient for $\sim$10,000~K). Given the fact that the current results agree above
300~K and that they were obtained with three independent theoretical approaches and four independent calculations, we
argue that the NLTE-ZDL and FGR-ZDL results are robust with an uncertainty
of less than 5\%. The LTE radiative association rate coefficients are also robust, and serve as upper limits corresponding to high density conditions. Numerical rate
coefficients are presented in Table~\ref{tab1}.

\begin{table}[h]
\centering
\caption{He$^+$ + H radiative association rate coefficients (10$^{-15}$ cm$^3$/s)\tablenotemark{a}} \label{tab1}
\begin{tabular}{lll}
\hline
\hline
\multicolumn{1}{c}{Temperature (K)} & \multicolumn{1}{c}{LTE} & \multicolumn{1}{c}{NLTE-ZDL}\\
\hline
10 & 4.125    & 1.383     \\
100 &  0.743   &  0.625     \\
1000 &  0.273   &   0.268    \\
2000 &  0.231   &   0.229   \\
3000 &  0.218  &   0.217 \\
4000 &  0.214    &  0.214      \\
5000 &  0.214    &  0.214 \\
7000 &  0.219    & 0.218 \\
10,000 & 0.231 & 0.230 \\
\hline
\end{tabular}
\tablenotetext{a}{Additional data are available in a machine-readable form in the online journal and on the website www.physast.uga.edu/ugamop/.}
\end{table}

Recently, \citet{vra13} studied the reverse process of photodissociation,
\begin{equation}
{\rm HeH}^+ + {\rm h}\nu \rightarrow {\rm He}^+ + {\rm H},
    \label{heh+pd}
\end{equation}
using a time-dependent approach. Using these results, they estimated the radiative association cross section~(\ref{he+hra}). However, their photodissociation calculations neglected the rotational dependence of the cross sections (i.e., only $J^{\prime\prime}=0$ was considered). As a consequence, they multiplied each $v^{\prime\prime}$ cross section by the total number of rotational states for that vibrational state. \citet{gus19} presented a temperature-independent rate coefficient, based on the above cross section, which we show in
Fig.~\ref{fig1}. This estimate is about a factor of 1.6
smaller than the current detailed calculations, which conversely sum over transitions to all $v^{\prime\prime},J^{\prime\prime}$ levels as indicated by
Eq.~(\ref{sigsum}). Hence, the approximation proposed in \citet{vra13} is not very reliable and, in fact, unnecessary.

\section{Discussion}

\citet{gus19} performed a chemical model of NGC 7027 near the Str\"omgren radius where the HeH$^+$ adundance is predicted to peak. Their model considers reactions~(\ref{he+hra}) and
(\ref{heh+rea}), as well as the dissociation recombination
process
\begin{equation}
    {\rm HeH}^+ + e \rightarrow {\rm He} + {\rm H}.
    \label{dr}
\end{equation}
They adopted updated rates for the latter two processes,
but found that use of the radiative association rate coefficient from \citet{vra13} resulted in a predicted abundance of a factor of $\sim$4 too small. It should be noted that they also performed a non-local thermodynamic equilibrium (NLTE) spectral model of the HeH$^+$ rotational emission assuming excitation by electron collisions. 
Their NLTE spectra model underpredicted the HeH$^+$ $J=1
\rightarrow 0$ rotational line intensity. To obtain agreement with observation, \citet{gus19} scaled the rate coefficient deduced from
\citet{vra13} by a factor of 4.3 as shown in Fig.~\ref{fig1}. They justified this arbitrary scaling by arguing that the radiative association rate coefficients in the literature vary widely. If one removes the approximate value deduced from the cross sections of \citet{vra13} and the highest temperature points from \citet{kra95}, which we believe are too low due to convergence issues, then there is a rather tight agreement between all previous calculations over the 1000-10,000~K range. The radiative association rate coefficient is robustly determined from prior data to be $(2.0\pm 0.5)\times 10^{-16}$ cm$^3$/s
in the relevant temperature range. The current set of calculations significantly reduce this uncertainty to $(2.14\pm 0.1)\times 10^{-16}$ cm$^3$/s at 5000 K (near the local minimum). We therefore argue that this arbitrary scaling of 4.3 (or now 2.8 if the current result is adopted) is unjustified, being at odds with statements given in \citet{gus19}. In fact, the uncertainties in the other two key reactions are probably larger. Further, scaling of rate coefficients to match observed intensities more likely hides neglected physics in the model.

For the case of dissociative recombination, a new storage ring measurement \citep{nov19} found a slight increase in the DR rate coefficient by a factor of $1.4\pm 0.3$ at
10,000~K. Unfortunately, this would have the affect of reducing the modeled HeH$^+$ abundance. 

Likewise, the NGC 7027 model of \citet{gus19} adopted the reactive scattering rate coefficients of \citet{bov11}, but comparison of their cross sections to earlier measurements suggest the scattering calculations are likely too large by a factor of $\sim$1.6 near 10$^4$~K. Such calculations involving a system of three atoms/ions are far more difficult to perform than two-body radiative association. Further, the time-dependent approach used by \citet{bov11} cannot distinguish the product channels as it incorporates a flux-absorbing imaginary potential. Instead their result is a sum over reactive and inelastic channels, a so-called depletion rate. Considering these reaction uncertainties suggests that the HeH$^+$ modeled abundance may be increased. On balance, a discrepancy factor between $\sim$3 to 4
remains. Similar conclusions were also reach by \citet{neu20}, who updated the model in \citet{gus19}.

One mechanism not considered in the current model is collisional excitation of HeH$^+$ due to H collisions; this process which might be relevant as the H/e$^-$ fractional abundance ratio is $\sim$10 just outside the Str\"omgren radius, where the abundance of HeH$^+$ peaks. Unfortunately, we are unaware of any rotational inelastic collisional rate coefficients of HeH$^+$ due to heavy particles. If one assumes an upper limit due to the HeH$^+$ + H depletion rate of 1.2$\times 10^{-9}$ cm$^3$/s compared to the electron rate coefficient of 2.7$\times 10^{-7}$ cm$^3$/s at 10,000~K, the contribution to the line intensity due to H collisions is about $\sim$5\% of that due to electrons. The omission of this collisional process is therefore unlikely to reduce the aforementioned discrepancy.

The consideration of a rotationally- and/or vibrationally-resolved chemistry may resolve some of the discrepancy. However, the sparsity of quantum-state-resolved chemical rates has generally limited such studies. \citet{agu10} considered the affect of reactions involving vibrationally-excited H$_2$ in a number of environments, including NGC 7027, while \citet{wal18} studied a full rovibrationally-resolved chemistry for H$_2$ and H$_2^+$ in the early Universe. The former do find a significant enhancement in the abundance of CH$^+$ in their NGC 7027 model, but it is formed by the endothermic process
\begin{equation}
    {\rm H}_2 + {\rm C}^+ \rightarrow {\rm CH}^+ + {\rm H},
\end{equation}    
while reaction (\ref{heh+rea}) is extremely exothermic. While we are unaware of rovibrationally-resolved reaction data for process (\ref{heh+rea}), the rate coefficients are unlikely to be very sensitive to internal excitation for kinetic temperatures of 5000-10,000~K. In fact, the recent DR measurements of \citet{nov19} find the rate coefficients for reaction (\ref{dr}) to be nearly independent of rotational excitation for $T>800$~K. 

Finally, infrared (IR) pumping of excited HeH$^+$ rovibrational levels due to thermal dust emission may be important. NGC 7027 has a prominent dust peak near 30~$\mu$m. IR radiative pumping may enhance the $J=1$ population and should be considered in the excitation model. For the case of H$_2^+$ in the recombination era, \citet{wal18} find that its level populations are driven to a thermal distribution characterized by the temperature of the cosmic background radiation field.

In summary, the first detection of HeH$^+$ in an astronomical source by \citet{gus19} is an important advance for molecular astrophysics. It will revitalize interest in helium chemistry. However, we show in this study that of all the reactions relevant to HeH$^+$ chemistry in NGC 7027, the radiative association of He$^+$ and H is probably the most accurately known, while the recent measurements of \citet{nov19} have put our knowledge of HeH$^+$ DR on a much sounder footing. Through accurate calculations performed in this work, we have narrowed the uncertainty in the radiative association rate coefficient further, and argue that arbitrary scaling of its rate coefficient to bring observations and models into agreement is not justified, and in fact, may hide important neglected physics. While detailed astrophysical modeling of the HeH$^+$ emission from NGC 7027 is beyond the scope of this work, we propose possible improvements to the chemistry, many of which we are
explicitly exploring in a robust investigation of helium reactions in the early Universe \citep{cou20}.

\vskip 1pc

This work was partially supported by NSF Grant No. PHY-1806180 and NASA Grant NNX15AI61G.
ITAMP is supported by a grant from the NSF to Harvard University and the Smithsonian Astrophysical Observatory. We thank David Neufeld for helpful comments.



\end{document}